\documentclass[aps,prd,preprint,nofootinbib,11pt]{revtex4}
\usepackage{amsfonts}
\usepackage{mathrsfs}
\usepackage{graphicx}
\usepackage{amsmath}
\usepackage{amssymb}
\usepackage{subfigure}
\usepackage{epsfig}
\usepackage{graphicx}
\usepackage{color}
\newcommand{\bqa}{\begin{eqnarray}}
\newcommand{\eqa}{\end{eqnarray}}
\newcommand{\beq}{\begin{equation}}
\newcommand{\eeq}{\end{equation}}
\allowdisplaybreaks[1]
\graphicspath{{fig/}{dia/}} \DeclareGraphicsExtensions{.eps}
\begin{document}

\title{Gluonic Tetracharm Configuration of $X(6900)$\\[0.7cm]}

\author{Bing-Dong Wan$^1$ and Cong-Feng Qiao$^{1,2}$\footnote{qiaocf@ucas.ac.cn}\vspace{+3pt}}

\affiliation{$^1$ School of Physics, University of Chinese Academy of Science, Yuquan Road 19A, Beijing 10049 \\
$^2$ CAS Center for Excellence in Particle Physics, Beijing 10049, China}

\author{~\\~\\}

\begin{abstract}
\vspace{0.5cm}

Recently, a new hadronic structure at around $6.9$ GeV was observed in an LHCb experiment. From its limited yet known decay mode, one could still determine that it contains at least four charm quarks and hence belongs to the category of exotic state. This finding indicates for the first time the tetracharm exotic states and is therefore quite importance. In this letter, we propose a nature hybrid interpretation for the structure of $X(6900)$, i.e. in $[\bar{3}_c]_{c c}\otimes[8_c]_{G}\otimes[3_c]_{\bar{c} \bar{c}}$ configuration with $J^{PC}=0^{++}$, and by using the QCD Sum Rule technique we performed mass spectrum calculation. The results showed that the observed $X(6900)$ could be a gluonic tetracharm state, and some other structures may exist, e.g., one around $7.2$ GeV in the tetracharm hybrid configuration and with $J^{PC}=0^{-+}$. We also predict the tetrabottom hybrid states, leaving for future experiment.
\end{abstract}
\pacs{11.55.Hx, 12.38.Lg, 12.39.Mk} \maketitle
\newpage

The establishment of quark model (QM) in 1950s is a milestone in the exploration of micro world \cite{GellMann:1964nj,Zweig}. The spectroscopy of conventional meson and baryon in QM is being gradually confirmed though experiments and will be completed soon. With the development of technology in the new millennium, the emergence of the so-called exotic state such as $X(3872)$ has been reported \cite{Choi:2003ue}, and new states tend to appear more frequently. Presently, we have a set of exotic state candidates that are waiting to be  characterized; this situation is similar to the phase of "particle zoo" witnessed in the last century. Discovering more exotic states and exploring their properties are currently one of the most intriguing and important topics in particle physics, which may promote our understanding of quantum chromodynamics (QCD) and enrich our knowledge on hadron spectroscopy.

In the light hadron spectrum, as the spacings between various states are normally small, it is difficult to split the exotic states from conventional hadrons, except when the former possess some peculiar quantum numbers. In contrast, the exotic states in the heavy hadron spectrum may have relatively distinct signatures. Indeed, in recent years, a set of so-called charmonium-/bottomonium-like states XYZ have been observed in experiments \cite{Choi:2003ue, Aubert:2005rm, Belle:2011aa, Ablikim:2013mio, Liu:2013dau}, which provides a new horizon for our understanding of the emergent structures in QCD.

Recently, in proton-proton collision at the center-of-mass energies of $\sqrt{s}=7$, 8 and 13 TeV, LHCb Collaboration revealed a narrow structure in $J/\psi$-pair invariant mass of approximately $6.9$ GeV with  significance greater than $5\; \sigma$ \cite{Aaij:2020fnh}. In additional to the narrow $X(6900)$, a broad structure just above the double $J/\psi$ threshold and another one at around $7.2$ GeV were also reported. This is for the first time that clear structures in the $J/\psi$-pair mass spectrum were observed in the experiment. If $X(6900)$ is further confirmed to be a hadronic structure, a tetracharm state rather merely some kinematic effect, the new finding will be considered as a huge breakthrough in the exploration of hadron spectroscopy.

\begin{figure}
\includegraphics[width=3.3cm]{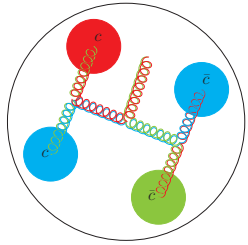}
\caption{A sketch of the tetracharm hybrid.} \label{cartoon}
\end{figure}

In the literature, some studies on the tetracharm(bottom) states have been conducted \cite{Iwasaki:1976cn,Chao:1980dv,Ader:1981db,Karliner:2016zzc,Barnea:2006sd,Debastiani:2017msn,Liu:2019zuc,Chen:2016jxd,Wang:2019rdo,Lloyd:2003yc,Anwar:2017toa,Wang:2017jtz,Jin:2020jfc,Lu:2020cns,Yang:2020rih,Wang:2020ols,Albuquerque:2020hio,Sonnenschein:2020nwn,Giron:2020wpx,liu:2020eha,Gong:2020bmg,Weng:2020jao,Dong:2020nwy,Yang:2020atz,Wang:2020wrp,Czarnecki-Leng-Voloshin}, and most of these theoretical studies predict resonants with masses in the range of the broad structure \cite{Aaij:2020fnh}. Of the narrow structure $X(6900)$, whose mass is higher than the double $J/\psi$ threshold by approximately $700$ MeV, it can be naturally attributed to certain excitation of the dicharm pair ground state. In this work, we conjecture that this structure is a tetracharm hybrid state, a kind of dicharm pair excitation containing a pair of constituent diquarks and a dynamic gluon, as shown in Fig. \ref{cartoon}.

In the gluonic tetracharm model, the $cc$ and $\bar{c}\bar{c}$ diquark pair lie in configurations of color $\bar{3}$ and (3) in the SU(3) gauge group respectively, which are relatively compact and were once used to interpret the pentaquak state $\Theta$(1540) \cite{Jaffe-Wilczek}. Hybrid is a kind of hadronic structure, which is accessible in light of QCD formalism, while has still no clear evidence in experiment. The hybrid charmonium model was applied to explain the Y(4260) \cite{Close-Page}, a somehow established exotic state.

We evaluate the four-quark hybrid by using the model with the independent Shifman, Vainshtein and Zakharov (SVZ) sum rule technique \cite{Shifman}. The SVZ sum rule, viz the QCD sum rule(QCDSR), has some peculiar advantages in exploring hadron properties involving nonperturbative QCD. It is a QCD based theoretical framework that incorporates nonperturbative effects universally order by order, rather a phenomenological model, and has already achieved much in the study of hadron spectroscopy and decays. To establish the sum rules, the starting point is to construct proper interpolating currents corresponding to the hadron of interest. By using the current, one can then construct the two-point correlation function, which has two representations: the QCD representation and the phenomenological representation.  Then, by equating these two representations, the QCDSR will be formally established, from which the hadron mass and decay width may be deduced.

In this work, to understand the nature of $X(6900)$, we investigated the gluonic tetracharm state in the QCD sum rule with composite current quantum numbers of $J^P=0^{++}$ and $0^{-+}$, which exhibit to be the simplest ones. Their bottom partners are also evaluated similarly.

The lowest order possible interpolating currents for $0^{++}$ and $0^{-+}$ tetracharm hybrid states take the following forms:
\begin{eqnarray}
j^{A}_{0^{++}} (x)&=& g_s \epsilon_{ikl}\epsilon_{jmn} [c_k^T C \gamma_\mu c_l]\frac{\lambda_{ij}^a}{2} G_{\mu\nu}^a [\bar{c}_m \gamma_\nu C \bar{c}_n^T]\, , \label{current-0++-A} \\
j^{B}_{0^{++}}(x) &=& g_s \epsilon_{ikl}\epsilon_{jmn} [c_k^T C \gamma_\mu\gamma_5 c_l]\frac{\lambda_{ij}^a}{2} G_{\mu\nu}^a [\bar{c}_m \gamma_\nu\gamma_5 C \bar{c}_n^T]\, , \label{current-0++-B} \\
j^{A}_{0^{-+}} (x)&=& g_s \epsilon_{ikl}\epsilon_{jmn} [c_k^T C \gamma_\mu c_l]\frac{\lambda_{ij}^a}{2} \tilde{G}_{\mu\nu}^a [\bar{c}_m \gamma_\nu C \bar{c}_n^T]\, , \label{current-0-+-A} \\
j^{B}_{0^{-+}}(x) &=& g_s \epsilon_{ikl}\epsilon_{jmn} [c_k^T C \gamma_\mu\gamma_5 c_l]\frac{\lambda_{ij}^a}{2} \tilde{G}_{\mu\nu}^a [\bar{c}_m \gamma_\nu\gamma_5 C \bar{c}_n^T]\, . \label{current-0-+-B}
\end{eqnarray}
Here, $g_s$ is the strong coupling constant, $i$, $j$, $k$, $\cdots$ are color indices, $\mu$ and $\nu$ are Lorentz indices, $\lambda^a$ are the Gell-Mann matrices, $C$ represents the charge conjugation matrix, $G_{\mu\nu}^a$ is the gluon field strength, and $\tilde{G}_{\mu\nu}^a=\frac{1}{2}\epsilon_{\mu\nu\alpha\beta} G^{a,\;\alpha\beta}$ denotes its dual field strength.

\begin{figure}
\includegraphics[width=6.8cm]{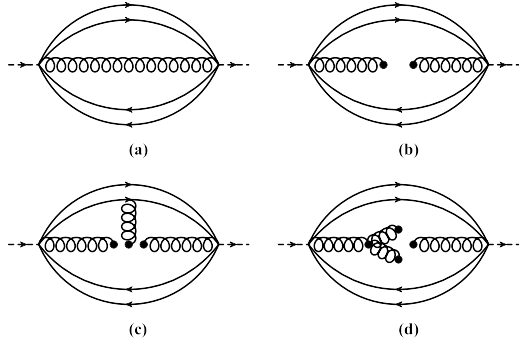}
\caption{The typical Feynman diagrams of a tetraquark hybrid state, where the permutation diagrams are implied. (a), (b), (c), and (d) represent respectively the contributions from perturbative, two-gluon condensate, and trigluon condensates.} \label{figfeyn}
\end{figure}

With the currents (\ref{current-0++-A})$-$(\ref{current-0-+-B}), the two-point correlation function can be readily established, i.e.,
\begin{eqnarray}
\Pi^{k}_{J^{PC}}(q^2) &=& i \int d^4 x\; e^{i q \cdot x} \langle 0 | T {\{} j^{k}_{J^{PC}}(x), j^{k}_{J^{PC}}(0)^\dagger {\}} |0 \rangle  \;,
\end{eqnarray}
where the subscript $J^{PC}$ represent for the quantum number of the involved hybrid state, $k$ runs from $A$ to $B$, and $|0\rangle$ denotes the physical vacuum. The Feynman diagrams of the correlation function in calculation are shown in Fig. \ref{figfeyn}.

On the QCD side, the correlation function $\Pi^{k}_{J^{PC}}$ can be expressed as a dispersion relationship:
\begin{eqnarray}
\Pi^{k,\;OPE}_{J^{PC}} (q^2) &=& \int_{s_{min}}^{\infty} d s
\frac{\rho^{k,\;OPE}_{J^{PC}} (s)}{s - q^2} \; .
\label{OPE-hadron}
\end{eqnarray}
Here, $\rho^{k,\;OPE}_{J^{PC}}(s) = \text{Im} [\Pi^{k,\;OPE}_{J^{PC}}(s)] / \pi$ and $s_{min}$ is a kinematic limit, which usually corresponds to the square of the sum of the current quark masses of the hadron \cite{Albuquerque:2013ija}, i.e., $s_{min}=16 m_c^2$. By applying the Borel transformation to (\ref{OPE-hadron}), we have the following equation:
\begin{eqnarray}
\Pi^{k,\;OPE}_{J^{PC}}\!\! = \!\!\int_{s_{min}}^{\infty} d s
\rho^{k,\;OPE}_{J^{PC}}(s)e^{- s / M_B^2}\ . \label{quark-gluon}
\end{eqnarray}

On the hadron side, after separating the ground state contribution from the pole terms, the correlation
function $\Pi^{k}_{J^{PC}}(q^2)$ is obtained as a dispersion integral over a physical regime, i.e.,
\begin{eqnarray}
\Pi^{k,\;phen}_{J^{PC}}(q^2) & = & \frac{(\lambda^{k}_{J^{PC}})^2}{(m^{k}_{J^{PC}})^2 - q^2} + \frac{1}{\pi} \int_{s_0}^\infty d s \frac{\rho^{k}_{J^{PC}}(s)}{s - q^2} \; , \label{hadron}
\end{eqnarray}
where  $m^{k}_{J^{PC}}$ denotes the mass of the lowest lying hybrid state, $\rho^{k}_{J^{PC}}(s)$ is the spectral density that contains the contributions from the higher excited states, $\lambda^{k}_{J^{PC}}$ is the coupling constant and the continuum states above the threshold $s_0$.

By performing the Borel transform on the hadronic side, Eq.(\ref{hadron}), and matching it to Eq.(\ref{quark-gluon}), we can then obtain the mass of the tetraquark hybrid state,
\begin{eqnarray}
m^{k}_{J^{PC}}(s_0, M_B^2) &=& \sqrt{-\frac{L^{k}_{J^{PC},\;1}(s_0, M_B^2)}{L^{k}_{J^{PC},\;0}(s_0, M_B^2)}} \; . \label{mass-Eq}
\end{eqnarray}
Here the moments $L_1$ and $L_0$ are defined as follows:
\begin{eqnarray}
L^{k}_{J^{PC},\;0}(s_0, M_B^2) & = & \int_{s_{min}}^{s_0} d s \; \rho^{k,\;OPE}_{J^{PC}}(s) e^{-s / M_B^2} \; , \label{L0}\\
L^{k}_{J^{PC},\;1}(s_0, M_B^2) & = & \frac{\partial}{\partial{\frac{1}{M_B^2}}}{L^{k,\;OPE}_{J^{PC}}(s_0, M_B^2)} \; .
\end{eqnarray}

In the numerical calculation, the input parameters are taken from
\cite{Shifman, Albuquerque:2013ija, Reinders:1984sr, P.Col, Narison:1989aq}:
$m_c (m_c) = \overline{m}_c= (1.275 \pm 0.025)\; \text{GeV}$,
$m_b (m_b) = \overline{m}_b= (4.18 \pm 0.03)\; \text{GeV}$,
$\langle g_s^2 G^2 \rangle = 0.88 \; \text{GeV}^4$,
$\langle g_s^3 G^3\rangle = 0.045 \; \text{GeV}^6$, in which the $\overline{\text{MS}}$ running heavy quark masses are adopted. Furthermore, the leading order strong coupling constant
\begin{eqnarray}
\alpha_s(M_B^2)=\frac{4\pi}{\big(11-\frac{2}{3}n_f \big) \text{ln} \big(\frac{M_B^2}{\Lambda_{\text{QCD}}^2}\big)}
\end{eqnarray}
with $\Lambda_{\text{QCD}} = 300$ MeV is taken, and $n_f$, here 5, represents the number of active quarks.

Moreover, there exist two additional parameters $M_B^2$ and $s_0$ that are introduced in establishing the sum rules, which will be fixed in light of the so-called standard procedures by fulfilling two criteria \cite{Shifman, Reinders:1984sr, P.Col,Albuquerque:2013ija}. The first one requires the convergence of the OPE. That is, one needs to compare individual contributions with the overall magnitude on the OPE side, and choose a reliable region for $M_B^2$ to retain the convergence. The second criterion requires that the portion of the lowest lying pole contribution (PC), the ground state contribution, in the total, pole plus continuum, should be over 50\%~\cite{Qiao:2014vva,Qiao:2015iea}. The two criteria can be formulated as follows:
\begin{eqnarray}
  R^{k,\;OPE}_{J^{PC}} = \left| \frac{L^{k,\;\langle g_s^3 G^3\rangle}_{J^{PC},\;0}(s_0, M_B^2)}{L^{k}_{J^{PC},\;0}(s_0, M_B^2)}\right|\, ,\label{RatioOPE}
\end{eqnarray}
\begin{eqnarray}
  R^{k,\;PC}_{J^{PC}} = \frac{L^{k}_{J^{PC},\;0}(s_0, M_B^2)}{L^{k}_{J^{PC},\;0}(\infty, M_B^2)} \; . \label{RatioPC}
\end{eqnarray}

To find a proper value for continuum threshold $s_0$, we perform a similar analysis as given in Refs.~\cite{Qiao:2014vva,Qiao:2015iea}. Therein, one needs to determine the proper value, which has an optimal window for the mass curve of the interested hadron. Within this window, the physical quantity, that is the mass of the concerned hadron, is independent of the Borel parameter $M_B^2$ as much as possible. In practice, we may vary $\sqrt{s_0}$ by $0.1$ GeV in numerical calculation \cite{Qiao:2014vva,Qiao:2015iea}, which sets the upper and lower bounds and hence the uncertainties of $\sqrt{s_0}$.

\begin{figure}
\includegraphics[width=6.8cm]{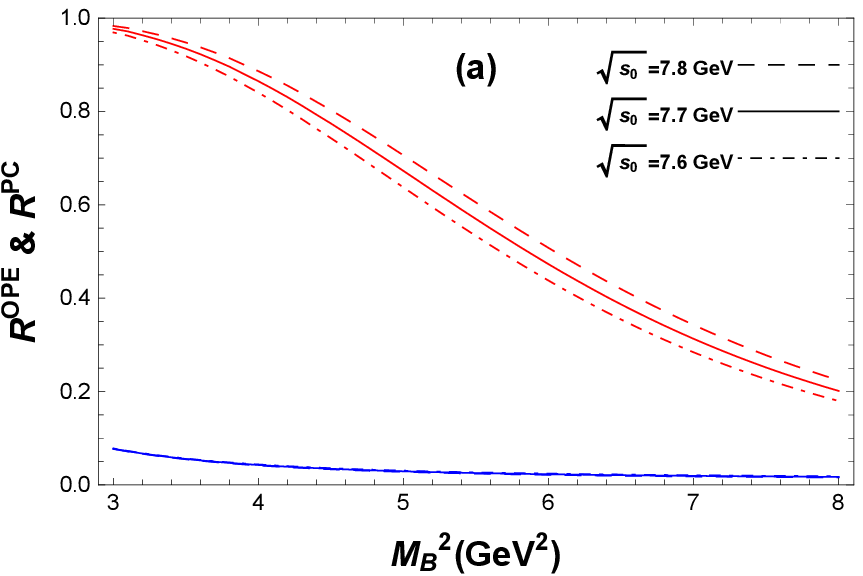}
\includegraphics[width=6.8cm]{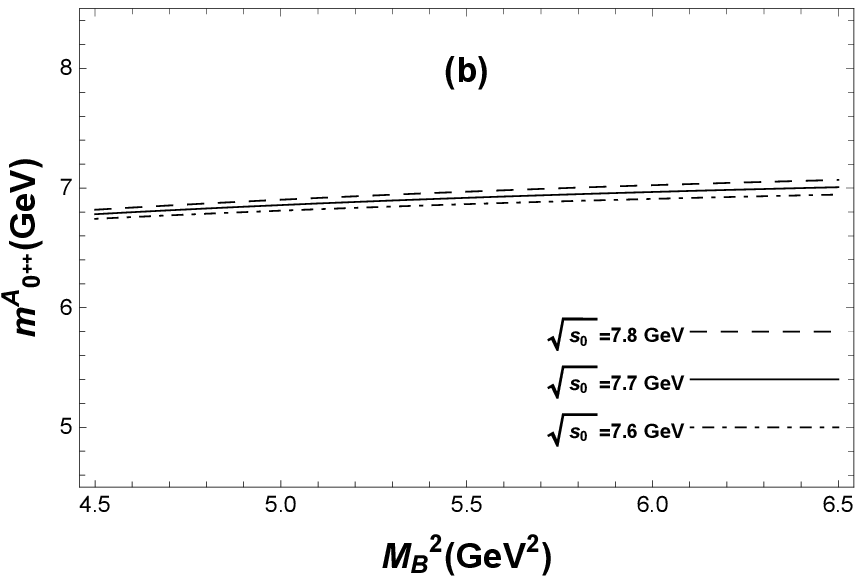}
\caption{ (a) The ratios ${R_{0^{++}}^{A,\;OPE}}$ and ${R_{0^{++}}^{A,\;PC}}$ as functions of the Borel parameter $M_B^2$ for different values of $\sqrt{s_0}$, where blue lines represent ${R_{0^{++}}^{A,\;OPE}}$ and red lines denote ${R_{0^{++}}^{A,\;PC}}$ . (b) The mass $M_{0^{++}}^{A}$ as a function of the Borel parameter $M_B^2$ for different values of $\sqrt{s_0}$.} \label{fig0++}
\end{figure}

\begin{figure}
\includegraphics[width=6.8cm]{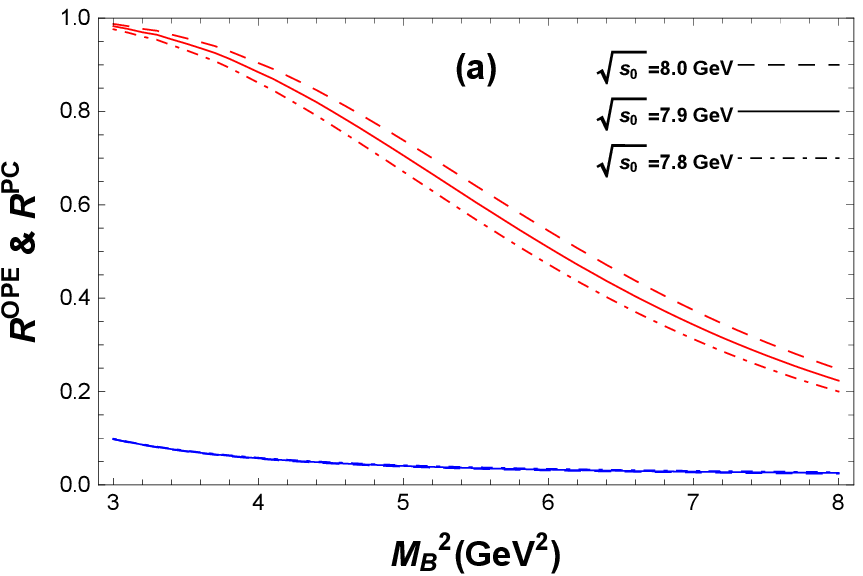}
\includegraphics[width=6.8cm]{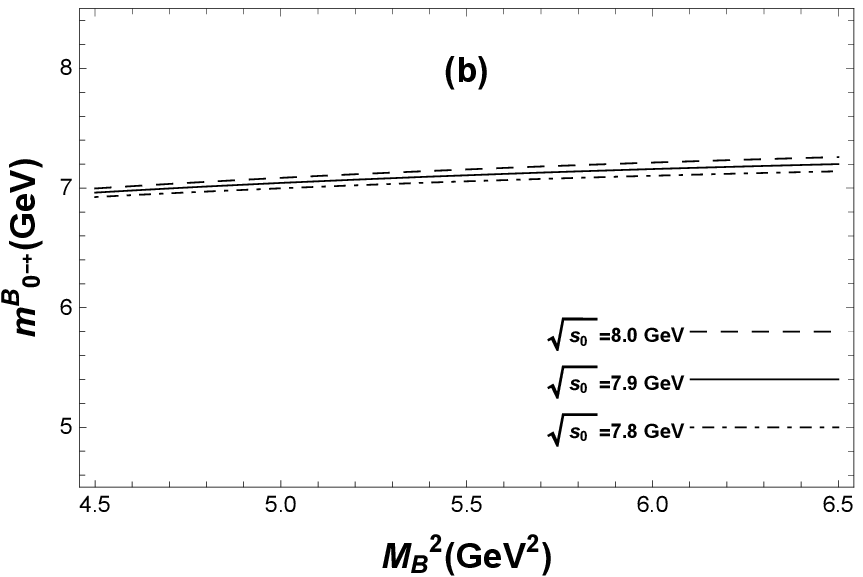}
\caption{The same caption as in Fig \ref{fig0++}, but for the ${R_{0^{-+}}^{B,\;OPE}}$, ${R_{0^{-+}}^{B,\;PC}}$, and $M_{0^{-+}}^{B}$.} \label{fig0-+}
\end{figure}

With the above preparation we numerically evaluate the mass spectrum of the tetracharm hybrid states. For the current in Eq. (\ref{current-0++-A}), the ratios ${R_{0^{++}}^{A,\;OPE}}$ and ${R_{0^{++}}^{A,\;PC}}$ are shown as functions of Borel parameter $M_B^2$ in Fig. \ref{fig0++}(a) with different values of $\sqrt{s_0}$, i.e., $7.6$, $7.7$, and $7.8$ GeV. The dependence relationships between $m_{0^{++}}^{A}$ and parameter $M_B^2$ are given in Fig. \ref{fig0++}(b). The optimal window of the Borel parameter is between $4.8 \le M_B^2 \le 5.9\; \rm{GeV}^2$, where a smooth section, the so called stable plateau, in the $m_{0^{++}}^{A}$-$M_B^2$ curve exists, which suggest the mass of the possible $0^{++}$ tetracharm hybrid state. The mass $m_{0^{++}}^{A}$ can be extracted as follows:
\begin{eqnarray}
&&m_{0^{++}}^{A} = (6.92 \pm 0.14) \, \text{GeV}  \; .\label{eq-mass-1}
\end{eqnarray}
This mass value is in good agreement with the observed mass of the $X(6900)$ state \cite{Aaij:2020fnh}, and implies a possible $J^{PC}=0^{++}$ hybrid configuration.

For currents (\ref{current-0++-B}) and (\ref{current-0-+-A}), we can not obtain positive spectral density functions $\rho^B_{0++}$ and$\rho^A_{0^{-+}}$, which implies that the current structures in Eqs. (\ref{current-0++-B}) and (\ref{current-0-+-A}) do not support the corresponding hybrid states.

For current (\ref{current-0-+-B}), the ratios ${R_{0^{-+}}^{B,\;OPE}}$ and ${R_{0^{-+}}^{B,\;PC}}$ as functions of the Borel parameter $M_B^2$ are shown in Fig. \ref{fig0-+}(a) with different values of $\sqrt{s_0}$ as well, and the relationship between $m_{0^{-+}}^{B}$ and the parameter $M_B^2$ is given in Fig. \ref{fig0-+}(b). The optimal window for the Borel parameter is found between $4.9 \le M_B^2 \le 6.0\; \rm{GeV}^2$, where a stable plateau in the $m_{0^{-+}}^{B}$-$M_B^2$ curve emerges, suggesting the mass of a possible $0^{-+}$ tetracharm hybrid state to be
\begin{eqnarray}
&&m_{0^{-+}}^{B} = (7.10 \pm 0.12) \, \text{GeV}  \; .\label{eq-mass-2}
\end{eqnarray}
This mass fits well with the signature of a hadronic structure recently observed by LHCb at around $7.2\; \rm{GeV}$ \cite{Aaij:2020fnh}.

In the results (\ref{eq-mass-1}) and (\ref{eq-mass-2}), the errors stem from the uncertainties of the quark masses, the Borel parameter $M_B^2$ and the threshold parameter $\sqrt{s_0}$.

Similarly, we can evaluate the tetrabottom hybrid states. By using the obtained analytical results but with $m_c$ being replaced by $m_b$, the corresponding masses are readily obtained, that is
\begin{eqnarray}
&&m^{A,\;b}_{0^{++}}=(19.30\pm 0.23)\, \text{GeV}  \; ,\label{eq-mass-3}\\
&&m^{B,\;b}_{0^{-+}}=(19.46\pm 0.20)\, \text{GeV}  \; .\label{eq-mass-4}
\end{eqnarray}
Here, superscript $b$ denotes the $b$-sector hybrid.

In summary, the gluonic tetracharm configuration, i.e. $[\bar{3}_c]_{c c}\otimes[8_c]_{G}\otimes[3_c]_{\bar{c} \bar{c}}$, is proposed to interpret the hadronic structure $X(6900)$ recently observed in the LHCb experiment. Here, we explore the hybrid states with quantum numbers $J^{PC}=0^{++}$ and $0^{-+}$, the lowest energy states, in the frame work of the QCD sum rule. Two stable hybrid states are obtained with masses of approximately $6.92$ and $7.10$ GeV for $J^{PC}=0^{++}$ and $0^{-+}$, respectively, which fit well with the measurements considering of the errors in theory and experiment. Moreover, the $b$-sector partners are also evaluated, and we find two tetrabottom hybrid states exist with masses $(19.30\pm 0.23)$ and $(19.46\pm 0.20)$ GeV for $0^{++}$ and $0^{-+}$, respectively.

\vspace{0.7cm} {\bf Acknowledgments}

This work was supported in part by the National Natural Science Foundation of China(NSFC) under Grant Nos. 11975236 and 11635009.

\newpage
\begin{widetext}
\appendix

\section{The construction of the interpolating currents}

The parity of $G^a_{\mu\nu}(x)$:
\begin{eqnarray}
P [G^a_{\mu\nu}(x)] P^{-1} &=& P[\partial_\mu A_\nu^a(x)-\partial_\nu A_\mu^a(x) + g_s f^{abc} A_\mu^b(x) A_\nu^c(x) ]P^{-1}\nonumber\\
&=&(-1)^\mu (-1)^\nu \times [\partial_\mu A_\nu^a(x)-\partial_\nu A_\mu^a(x) + g_s f^{abc} A_\mu^b(x) A_\nu^c(x) ]\nonumber\\
&=&(-1)^\mu (-1)^\nu G^a_{\mu\nu}(x)\;,
\end{eqnarray}
with
\begin{eqnarray}
P[\partial^\mu]P^{-1}=(-1)^\mu \partial^\mu  \;,\; P[A^\mu (x)]P^{-1}=(-1)^\mu A^{\mu}(x).
\end{eqnarray}
Here, we use the shorthand $(-1)^\mu\equiv 1$ for $\mu=0$ and $(-1)^\mu\equiv -1$ for $\mu=1$, $2$, $3$.
Then the P-parity of $\tilde{G}^a_{\mu\nu}(x)$ can be deduced as follows:
\begin{eqnarray}
P [\tilde{G}^a_{\mu\nu}(x)] P^{-1} &=& P [\frac{1}{2}\epsilon_{\mu\nu\alpha\beta} G^{a,\;\alpha\beta}(x)] P^{-1}\nonumber\\
 &=&P[\frac{1}{2}\epsilon_{\mu\nu\alpha\beta}(\partial^\alpha A^{a\beta }(x)-\partial^\beta A^{a\alpha}(x) + g_s f^{abc} A^{b\alpha}(x) A^{c\beta}(x)) ]P^{-1}\nonumber\\
&=&(-1)^\alpha (-1)^\beta \times\frac{1}{2}\epsilon_{\mu\nu\alpha\beta}(\partial^\alpha A^{a\beta }(x)-\partial^\beta A^{a\alpha}(x) + g_s f^{abc} A^{b\alpha}(x) A^{c\beta}(x))\nonumber\\
&=&(-1)^\alpha (-1)^\beta \tilde{G}^a_{\mu\nu}(x)\;.
\end{eqnarray}
Because the subscripts $\mu$, $\nu$, $\alpha$ and $\beta$ are in the total antisymmetric factor $\epsilon_{\mu\nu\alpha\beta}$, it is not difficult to prove that $(-1)^\alpha(-1)^\beta=(-1)(-1)^\mu(-1)^\nu$. Then we have
\begin{eqnarray}
P [\tilde{G}^a_{\mu\nu}(x)] P^{-1} &=&(-1)(-1)^\mu(-1)^\nu\tilde{G}^a_{\mu\nu}(x)\;.
\end{eqnarray}

On the other hand, because
\begin{eqnarray}
 P[c^T C \gamma_\mu c] [\bar{c} \gamma_\nu C \bar{c}^T]P^{-1}&=&(-1)^\mu(-1)^
 \nu[c^T C \gamma_\mu c] [\bar{c} \gamma_\nu C \bar{c}^T]\;,\\
  P[c^T C \gamma_\mu\gamma_5 c] [\bar{c} \gamma_\nu\gamma_5 C \bar{c}^T]P^{-1}&=&(-1)^\mu(-1)^\nu[c^T C \gamma_\mu\gamma_5 c] [\bar{c} \gamma_\nu\gamma_5 C \bar{c}^T]\;,
 \end{eqnarray}
the $J^P$ of Eqs. (1) and (2) reads $0^{+}$, and the $J^P$ of Eqs. (3) and (4) is $0^{-}$.

 To maintain the invariance of the quark-gluon coupling under charge conjugate, i.e., $\hat{C}(\bar{\Psi} A^a_\mu T^a \gamma^\mu \Psi)\hat{C}^{-1} = \bar{\Psi} A^a_\mu T^a \gamma^\mu \Psi $, we need:
 \begin{eqnarray}
 [\hat{C} A^a_\mu T^a \hat{C}^{-1}] &=& - A^a_\mu (T^a)^{T} .
 \end{eqnarray}
 Here, $T^a=\frac{\lambda^a}{2}$. It can be proved that
 \begin{eqnarray}
 \hat{C}[f^{abc} A^{b}(x) A^{c}(x)] \hat{C}^{-1} &=& \text{sign}{\{}\hat{C}A^a(x)\hat{C}^{-1}{\}}f^{abc} A^{b}(x) A^{c}(x),
 \end{eqnarray}
by checking all possibilities of $\hat{C}[f^{abc} A^{b}(x) A^{c}(x)] \hat{C}^{-1}$ with certain magnitudes of $a$, $b$, and $c$, where $\text{sign}{\{}\hat{C}A^a(x)\hat{C}^{-1}{\}}$ denotes the C-parity of the gluon field $A^a(x)$.
Then, the C-parity of the gluon field strength $G^a_{\mu\nu}(x)$ can be deduced as follows:
\begin{eqnarray}
\hat{C} [G^a_{\mu\nu}(x)] \hat{C}^{-1} &=& \hat{C}[\partial_\mu A_\nu^a(x)-\partial_\nu A_\mu^a(x) + g_s f^{abc} A_\mu^b(x) A_\nu^c(x) ]\hat{C}^{-1}\nonumber\\
&=& \text{sign}{\{}\hat{C}A^a(x)\hat{C}^{-1}{\}}G^a_{\mu\nu}(x).
\end{eqnarray}
Similarly,
\begin{eqnarray}
\hat{C} [\tilde{G}^a_{\mu\nu}(x)] \hat{C}^{-1} &=&  \text{sign}{\{}\hat{C}A^a(x)\hat{C}^{-1}{\}}\tilde{G}^a_{\mu\nu}(x).
\end{eqnarray}
Therefore,
\begin{eqnarray}
\hat{C} [G^a_{\mu\nu}(x) T^a] \hat{C}^{-1} &=&  - G^a_{\mu\nu}(x) (T^a)^T\;,\label{Cparity_G}\\
\hat{C} [\tilde{G}^a_{\mu\nu}(x) T^a] \hat{C}^{-1} &=&  - \tilde{G}^a_{\mu\nu}(x) (T^a)^T\;. \label{Cparity_G1}
\end{eqnarray}

It is not difficult to prove that the Lorentz indices of $\mu$ and $\nu$ will be exchanged in charge conjugate transformation. Because $G^a_{\mu\nu}=-G^a_{\nu\mu}$, the minus signs in Eqs. (\ref{Cparity_G}) and (\ref{Cparity_G1}) will be canceled out. Similarly, the transposition signs $T$ in $T^a$ will be eliminated by the color indices exchange. Therefore, the C-parity of all the interpolating currents in Eqs. (\ref{current-0++-A}) to (\ref{current-0-+-B}) is $+$.

\section{The spectral densities of tetracharm hybrid}
To calculate the spectral density of the operator product expansion (OPE) side, the heavy-quark ($Q=c$ or $b$) full propagator $S^Q_{i j}(p)$ is used as follows:
\begin{eqnarray}
S^Q_{j k}(p) \! \! & = & \! \! \frac{i \delta_{j k}(p\!\!\!\slash + m_Q)}{p^2 - m_Q^2} - \frac{i}{4} \frac{t^a_{j k} G^a_{\alpha\beta} }{(p^2 - m_Q^2)^2} [\sigma^{\alpha \beta}
(p\!\!\!\slash + m_Q)
+ (p\!\!\!\slash + m_Q) \sigma^{\alpha \beta}]  \; .
\end{eqnarray}
Here, the vacuum condensates are explicitly shown. For more explanation on above propagators, the readers may refer to Refs.~\cite{Albuquerque:2013ija}.
In the QCD representation, the spectral density may be expressed as follows:
\begin{eqnarray}
\rho^{OPE}(s) & = & \rho^{pert}(s)  + \rho^{\langle G^2 \rangle}(s)  + \rho^{\langle G^3 \rangle}(s)\; . \label{rho-OPE}
\end{eqnarray}

\subsection{The spectral densities for $0^{++}$ tetracharm hybrid}
The spectral density $\rho^{OPE}(s)$ is calculated up to dimension six. For all currents shown in Eqs. (\ref{current-0++-A}) and (\ref{current-0++-B}), we obtain the spectral densities as follows:

\begin{eqnarray}
\rho_i^{pert}(s)&=&\frac{g_s^2}{2^{10}\times 5\times \pi^8} \int^{x_{+}}_{x_{-}} d x \int^{y_{+}}_{y_{-}} d y \int^{z_{+}}_{z_{-}} d z \int^{w_{+}}_{w_{-}} d w  A_{xyzw} H_{xyzw}^3\nonumber\\
&\times& \bigg{\{} 4 A_{xyzw} H_{xyzw}^3 x y z w  -2 H_{xyzw}^2 [18 xyzw A_{xyzw}s+{\cal N}_i(xy+zw)m_c^2]\nonumber\\
&+& 20 A_{xyzw} (m_c^4 - x y z w s^2) +5 H_{xyzw}[12 x y z w A_{xyzw}s^2\nonumber\\
&+&{\cal N}_i (x y+z w)m_c^2 s + 2 (A_{xyzw}-1)m_c^4]\bigg{\}}\;,\\
\rho_i^{\langle G^2 \rangle}(s)&=&\frac{\langle g_s^2 G^2\rangle}{2^{9}\times 3\times \pi^6} \int^{x_{+}}_{x_{-}} d x \int^{y_{+}}_{y_{-}} d y \int^{z_{+}}_{z_{-}} d z F_{xyz}^2\nonumber\\
&\times&\bigg{\{}6m_c^4+{\cal N}_i m_c^2(3 s-2 F_{xyz})(x y +z B_{xyz})\bigg{\}}\;,\\
\rho_i^{\langle G^3 \rangle}(s)&=&\frac{\langle g_s^3 G^3\rangle}{2^{8}\times \pi^6} \int^{x_{+}}_{x_{-}} d x \int^{y_{+}}_{y_{-}} d y \int^{z_{+}}_{z_{-}} d z \bigg{\{}\frac{1}{2}\big [4x y z B_{xyz} F_{xyz}^3\nonumber\\
&-&s(m_c^2+{\cal N}_i x y s)(m_c^2+{\cal N}_i z s B_{xyz})- 3F_{xyz}^2\big( 6s x y z B_{xyz}+ {\cal N}_i m_c^2(x y +z B_{xyz}) \big)\nonumber\\
&+&2F_{xyz}\big( m_c^4 +6s^2 x y z B_{xyz} +3 {\cal N}_i s m_c^2 (xy +z B_{xyz})  \big) \big] \nonumber\\
&+&\frac{m_c^2 F_{xyz}}{x}\big[ {\cal N}_i(F_{xyz}-s)(x y + z B_{xyz}) -2 m_c^2  \big]\bigg{\}}\;,
\end{eqnarray}
where the subscript $i$ runs from $A$ to $B$, and the factor ${\cal N}_i$ has the following definition: ${\cal N}_A = 1$ and ${\cal N}_B = -1$. Here, we use the following definitions:

\begin{eqnarray}
&&A_{xyzw}=(1-x-y-z-w)\;,B_{xyz}=(1-x-y-z)\;,\\
&&H_{xyzw}=\bigg(\frac{1}{x}+\frac{1}{y}+\frac{1}{z}+\frac{1}{w}  \bigg)m_c^2-s\;,\\
&&F_{xyz}=\bigg(\frac{1}{x}+\frac{1}{y}+\frac{1}{z}+\frac{1}{1-x-y-z}  \bigg)m_c^2-s\;,\\
&&x_{\pm}=\bigg[\bigg( 1-\frac{8m_c^2}{s} \bigg) \pm \sqrt{\bigg( 1-\frac{8m_c^2}{s} \bigg)^2-\frac{4m_c^2}{s}}\bigg] \bigg/2\;,\\
&&y_{\pm}=\bigg[ 1+2 x +\frac{3 s x^2}{m_c^2-s x} \pm \sqrt{\frac{[m_c^2+s x (x-1)][(8x+1)m_c^2+s x(x-1)]}{(m_c^2-s x)^2}}  \bigg] \bigg/2\;,\\
&&z_{\pm}=\bigg[(1-x-y)\pm \sqrt{\frac{(x+y-1)[m_c^2(x+y-(x -y)^2)+s x y(x+y-1)]}{s x y-(x+y) m_c^2}} \bigg]\bigg/2\;,\\
&&w_{-}=\frac{x y z m_c^2}{s x y z -(x y +y z + x z)m_c^2}\;,w_{+}=1-x-y-z\;.
\end{eqnarray}

\subsection{The spectral densities for $0^{-+}$ tetracharm hybrid}
For all currents given in (\ref{current-0-+-A}) and (\ref{current-0-+-B}), we obtain the spectral densities as follows:

\begin{eqnarray}
\rho_i^{pert}(s)&=&\frac{g_s^2}{2^{10}\times 5\times \pi^8} \int^{x_{+}}_{x_{-}} d x \int^{y_{+}}_{y_{-}} d y \int^{z_{+}}_{z_{-}} d z \int^{w_{+}}_{w_{-}} d w  A_{xyzw} H_{xyzw}^3\nonumber\\
&\times& \bigg{\{} -4 A_{xyzw} H_{xyzw}^3 x y z w  + 2 H_{xyzw}^2 [18 xyzw A_{xyzw}s+{\cal N}_i (3A_{xyzw}-1)\nonumber\\
&\times&(xy+zw)m_c^2]+ 20 s A_{xyzw} (m_c^2 +{\cal N}_i x y s) (m_c^2 +{\cal N}_i w z s)\nonumber\\
&-&5 H_{xyzw}[12 x y z w A_{xyzw}s^2+{\cal N}_i  (6 A_{xyzw}-1) (xy+zw)m_c^2 s\nonumber\\
&+& 2 (A_{xyzw}-1)m_c^4]\bigg{\}}\;,\\
\rho_i^{\langle G^2 \rangle}(s)&=&\frac{\langle g_s^2 G^2\rangle}{2^{9}\times 3\times \pi^6} \int^{x_{+}}_{x_{-}} d x \int^{y_{+}}_{y_{-}} d y \int^{z_{+}}_{z_{-}} d z F_{xyz}^2\nonumber\\
&\times&\bigg{\{}6m_c^4+{\cal N}_i m_c^2(3 s-2 F_{xyz})(x y +z B_{xyz})\bigg{\}}\;,\\
\rho_i^{\langle G^3 \rangle}(s)&=&\frac{\langle g_s^3 G^3\rangle}{2^{8}\times \pi^6} \int^{x_{+}}_{x_{-}} d x \int^{y_{+}}_{y_{-}} d y \int^{z_{+}}_{z_{-}} d z \bigg{\{}-\frac{1}{2}\big [4x y z B_{xyz} F_{xyz}^3\nonumber\\
&+&s(m_c^4-s^2 x y z B_{xyz})- 18 F_{xyz}^2 s x y z B_{xyz}+2 F_{xyz}(6 s^2 x y z B_{xyz} - m_c^4)  \big] \nonumber\\
&+&\frac{m_c^2 F_{xyz}}{x}\big[ {\cal N}_i(F_{xyz}-s)(x y + z B_{xyz}) -2 m_c^2  \big]\bigg{\}}\;.
\end{eqnarray}

\end{widetext}
\end{document}